\documentclass{article}
\usepackage{arxiv}

\usepackage[utf8]{inputenc} 
\usepackage[T1]{fontenc}    
\usepackage{hyperref}       
\usepackage{url}            
\usepackage{booktabs}       
\usepackage{amsfonts}       
\usepackage{nicefrac}       
\usepackage{microtype}      
\usepackage{lipsum}
\usepackage{graphicx}
\graphicspath{ {./images/} }
\usepackage{algpseudocode}
\usepackage[linesnumbered,ruled,vlined]{algorithm2e}
\usepackage{amsmath,amssymb,amsfonts}
\usepackage{graphicx}
\usepackage{textcomp}
\usepackage{xcolor}
\usepackage{cite}
\usepackage{pifont}
\usepackage{multirow}
\usepackage[shortlabels]{enumitem}
\usepackage{tabularray}
\usepackage{color,soul}
\usepackage{caption}
\usepackage{subcaption}
\usepackage[numbers]{natbib}
\title{D-CAPTCHA++: A Study of Resilience of Deepfake CAPTCHA under Transferable Imperceptible Adversarial Attack}

\author{
 Hong-Hanh Nguyen-Le \\
  School of Computer Science\\
  University College Dublin\\
  Dublin 4, Ireland \\
  \texttt{hong-hanh.nguyen-le@ucdconnect.ie} \\
   \And
 Van-Tuan Tran \\
  School of Computer Science and Statistics\\
  Trinity College Dublin\\
  Dublin 2, Ireland \\
  \texttt{tranva@tcd.ie} \\
  \And
 Dinh-Thuc Nguyen \\
  University of Science\\
  Ho Chi Minh City, Vietnam\\
  \texttt{ndthuc@fit.hcmus.edu.vn} \\
 \And
  Nhien-An Le-Khac\\
  School of Computer Science\\
  University College Dublin\\
  Dublin 4, Ireland \\
  \texttt{an.lekhac@ucd.ie} \\
}

\begin{document}
\maketitle
\begin{abstract}
The advancements in generative AI have enabled the
improvement of audio synthesis models, including text-to-speech and voice conversion. This raises concerns about its potential misuse in social manipulation and political interference, as synthetic speech has become indistinguishable from natural human speech. Several speech-generation programs are utilized for malicious
purposes, especially impersonating individuals through phone calls. Therefore, detecting fake audio is crucial to maintain social security and safeguard the integrity of information. Recent
research has proposed a D-CAPTCHA system based on the challenge-response protocol to differentiate fake phone calls from real ones. In this work, we study the resilience of this system and introduce a more robust version, D-CAPTCHA++, to defend
against fake calls. Specifically, we first expose the vulnerability of the D-CAPTCHA system under transferable imperceptible adversarial attack. Secondly, we mitigate such vulnerability by improving the robustness of the system by using adversarial training in D-CAPTCHA deepfake detectors and task classifiers.
\end{abstract}


\section{Introduction}
Over the past few years, audio synthesis models \cite{baas2023voice}, including text-to-speech (TTS) and voice conversion (VC), have significantly improved quality.
%
%
These technologies aim at generating more believable human-like natural speech with high quality and fast inference, which leads to the difficulty in distinguishing between real audio and fake ones.
Moreover, the accessibility of these technologies for creating and distributing spoofed speech has become easier through the internet and open resources.
The combination of developments in believability and accessibility to deepfake voice has posed serious threats to social security and the political economy, including impersonation, and voice cloning for fake phone/video calls \cite{Verma_2023, Zhadan_2023}.
Due to the ever-growing threat of fake audio, developing a reliable detection technique is imperative.

%
%

%
%
One of the detection approaches aims to cast the deepfake detection problem as a classification problem that learns a hard decision boundary to separate synthetic audio from human audio through a single deep learning (DL) model \cite{tak2021end, kawa2022specrnet, kawa2022defense}.
%
%
By contrast, a few studies introduced a challenge-response protocol that requires the user to respond to a challenge within a limited time. 
These methods are based on the assumption that the challenges can be difficult for artificial intelligence (AI) systems to understand while being easily performed by humans. 
Google's reCaptcha \cite{reCAPTCHA} and NLP Captcha \cite{NLPCAPTCHA} are early presented as a potential defense against hidden voice commands.
%
Recently, as an extension of these methods, \citeauthor{yasur2023deepfake} \cite{yasur2023deepfake} introduced a deepfake CAPTCHA (D-CAPTCHA) system for detecting fake calls.
This multifaceted system integrates numerous modules, encompassing Human-based, Time, Realism, Task, and Identity, to ensure AI systems produce speech responses within a limited time while maintaining human-like natural speech.
This system is designed based on two hypotheses: (1) Challenges cannot be understood easily by AI systems; (2) VC systems are unable to create speech content in real-time. Moreover, VC systems need to be retrained if the challenge is out of domain.
%

%
In this research, we focus on investigating the resilience of this complicated and effective system by proposing a transferable imperceptible adversarial attack method under the black-box scenario and proposing a method for improving the robustness of this system.
Our attack method is grounded on key hypotheses: (1) The challenge posed by the system can be comprehended and executed by the attacker; (2) A state-of-the-art VC model can generate synthetic speech in a limited time; (3) Imperceptible adversarial examples not only fool deepfake detectors but also preserve the semantic content that can bypass the Task module.
After empirical experiments, we expose the vulnerability of the D-CAPTCHA system, especially the interconnection between the deepfake detector and task classification modules, to the transferable imperceptible adversarial samples.
To mitigate the vulnerability of the D-CAPTCHA system, we introduce a more robust version of D-CAPTCHA, D-CAPTCHA++, by employing Projected Gradient Descent (PGD) adversarial training.
Our empirical experiments show that D-CAPTCHA++ can reduce the success rate of the transferable adversarial attacks from $31.31 \% \pm 1.40$ to $0.60 \% \pm 0.09$ for the task classifier and from $32.26\% \pm 0.99$ to $2.27 \% \pm 0.18$ for the deepfake detector, respectively.
Our work contributes to a more robust defense method against fake phone calls.

\textbf{Contributions.} The main contributions of this work are as follows:

\begin{itemize}
    \item We propose a semi-automated threat model that leverages the power of synthesis models and adversarial example-generation techniques.
    \item We introduce a simple yet effective mitigation, adversarial
training, to improve the robustness of the D-CAPTCHA system.
    \item We conduct extensive experiments on various voice conversion models (kNN-VC, Urhythmic, TriAAN), deepfake detectors (SpecRNet, RawNet2, RawNet3), and task classifiers (ResNet, RawNet3) to evaluate our proposed method.
    \item We analyze the impact of feature extraction techniques on imperceptible adversarial examples, which contributes to limiting the adversarial transferability in designing voice-based deepfake detection systems.
\end{itemize}
The subsequent sections of this paper are structured as follows. 
After this introduction, the background knowledge associated with the D-CAPTCHA system, voice conversion, and adversarial example generation techniques is reviewed in Section \ref{sec:bg}. 
In Section \ref{sec:method}, our methodology is described while experimental settings and results are presented in Section \ref{sec:experiments}.
Finally, Section \ref{sec:discussion} gives the conclusion and future work.

\section{Background}\label{sec:bg}
\subsection{Deepfake-CAPTCHA (D-CAPTCHA) System}\label{ssec:d-captcha}
D-CAPTCHA \cite{yasur2023deepfake} is a defense system to deepfake calls through a challenge-response protocol, which is designed based on two hypotheses: (i) Challenges cannot be understood easily by AI systems; (ii) VC systems are unable to create speech content involving the assigned challenge if it is out of trained domain.
The system includes five modules:
\begin{itemize}
    \item \textbf{Human-based} $\mathcal{H}$ module: Upon receiving a call from an unknown caller, the victim initially assesses whether the call raises suspicion. If so, the system will automatically record a voice sample $a_0$ from the caller. Subsequently, a random challenge, denoted as $c$, is assigned to the caller, who is then tasked with providing a corresponding response $r_c$.

    \item \textbf{Time} $\mathcal{T}$ module: This module constraints the caller to respond to the challenge $c$ in a limited time, particularly in $1s$. 

    \item \textbf{Realism} $\mathcal{R}$ module: $\mathcal{R}$ can be deepfake detectors that verify whether $r_c$ is spoofed or real. 

    \item \textbf{Task} $\mathcal{C}$ module: The goal of this module is to guarantee $r_c$ contains the requested task which is determined based on ML classifiers.

    \item \textbf{Identity} $\mathcal{I}$ module: $\mathcal{I}$ prevents the attacker from changing the identity during the challenge. The module evaluates the similarity between $a_0$ and $r_c$.
\end{itemize}

D-CAPTCHA exhibits a sophisticated design, making it exceedingly challenging for adversaries to successfully initiate fraudulent calls.
Moreover, the list of challenges can be extensive, further hindering the attacker's ability to bypass the system.
Specifically, the adversary must contrive a human-like response $r_c$ that not only deceives $\mathcal{R}$ module but also contains the challenge content required to bypass the $\mathcal{C}$ module, all while preserving the caller's identity.
However, there exist three main limitations of this system: 
\begin{enumerate}[(i)]
    \item $\mathcal{R}$ module is vulnerable to adversarial examples and can be evaded by adding a crafted perturbation to the response $r_c$;
    \item The main limitation of the $\mathcal{C}$ module lies in its inability to understand the semantic content of the response $r_c$;
    \item $\mathcal{I}$ module only compares between $a_0$ and $r_c$, leaving it vulnerable to circumvention by the adversary using a VC before and during the challenge period.
\end{enumerate}
Additionally, the current advancements in generative AI have enabled VC systems to produce high-quality audio in a remarkably short time \cite{baas2023voice, van2023rhythm, park2023triaan}.
Furthermore, the emergence of large language models (LLMs) \cite{zhao2023survey} like chatGPT, has opened up the possibility of AI systems comprehending challenge requirements shortly.
While our current work focuses on a semi-automated threat model where the attacker understands and executes the challenge using VC, we acknowledge the potential of extending our threat model to include LLMs that can autonomously tackle the challenge (discussed in Section \ref{sec:discussion}).

\subsection{Voice Conversion (VC)}\label{ssec:vc}
The primary objective of a VC system is to modify the identity-specific attributes of a source speaker, including timbre, pitch, and rhythm while carrying over the linguistic content.
In general, the operation of a VC system involves two phases: the training phase and the conversion phase. 
During the training phase, vocal data is extracted from both source and target speech to develop a conversion function, represented as $F$.
In the conversion step, given a source speech signal $x$, a corresponding feature representation $z$ is extracted from $x$ using a feature extraction $A$. This extracted feature representation $z$ is then passed through the conversion function $F$ that manipulates the source speech characteristics to align with those of the target speech. 
Finally, an inverse function $R$ is applied to convert the modified feature representation into an audible speech signal.
Formulatively, the flow of a voice conversion system can be represented as:
\begin{equation*}
    y = (R \circ F \circ A)(x)
\end{equation*}
, where $y$ is the target speech \cite{sisman2020overview}.

\begin{table*}[!ht]
\centering
\fontsize{10pt}{12pt}\selectfont
\caption{Threat model of the proposed attack}
\label{tab:threat-model}
\begin{tabular}{c|c|c}
\textbf{Threat Model Characteristics}  & \textbf{Type}                                                         & \textbf{Attacker View} \\ \hline
\multirow{9}{*}{Attacker's Knowledge}  & Task                                                                  & \ding{51}                     \\
                                       & Training data                                                         & \ding{55}                      \\
                                       & Preprocessing                                                         & \ding{55}                      \\
                                       & Feature Extraction                                                    & \ding{55}                       \\
                                       & Model's Architecture                                                  & \ding{55}                       \\
                                       & Objective function                                                    & \ding{55}                       \\
                                       & Model's Parameters                                                    & \ding{55}                       \\
                                       & Inference API                                                         & \ding{55}                       \\
                                       & Model's Output                                                        & \ding{51}                      \\ \hline
\multirow{2}{*}{Attacker's Goal}       & Integrity violation                                                               & \ding{51}                       \\
                                       & Availability violation                                                             & \ding{55}                       \\ \hline
\multirow{3}{*}{Attacker's Capability} & Manipulate training data                                              & \ding{55}                       \\
                                       & Manipulate test data                                                  & \ding{51}                       \\
                                       & Manipulate model                                                      & \ding{55}                       \\ \hline
\multirow{3}{*}{Attacker's Strategy}   & \multicolumn{1}{l|}{Train a surrogate model for parameter extraction} & \ding{55}                       \\
                                       & \multicolumn{1}{l|}{Train a surrogate model for transferability}      & \ding{51}                       \\
                                       & \multicolumn{1}{l|}{Generate imperceptible adversarial samples}       & \ding{51}                       \\ \hline
\end{tabular}
\end{table*}

\subsection{Adversarial Example Generation}\label{ssec:adversarial}
Adversarial examples are generated by adding a crafted perturbation to an input sample to make a classifier misbehave, which is considered an adversarial attack.
Given a classifier $f$ and some input-label pair $(x,y)$, the objective of an adversarial attack is to find a perturbation $\delta$ that alters the $f$'s decision for a perturbed input $x_{adv} = x + \delta$ by minimizing the $f$'s classification certainty:
\begin{equation}
    \text{min} \; \mathcal{L}(f(x + \delta), y) \; \text{subject to} \; ||\delta|| < \epsilon
    \label{eq:adversarial}
\end{equation}
, where $\mathcal{L}(f(x), y)$ is a loss function that is minimized when $f(x) = y$, $\epsilon$ is a hyperparameter to control the maximum perturbation, $||.||$ is the max-norm of $\delta$.
Depending on the assumptions about the adversary's knowledge of the target model, adversarial attacks can be classified as white-box attacks and black-box attacks.
In the white-box attack setting, attackers are assumed to have full knowledge of the target model $f$, including model parameters, architecture, training data, and thresholds.
In this scenario, the adversary performs gradient descent on the loss function $\mathcal{L}$ to generate adversarial samples.
However, real-world applications often deploy models as APIs, limiting the knowledge and access of the attacker to the model.
This scenario is regarded as a black-box attack, where the attacker is not allowed to analytically compute the gradient descent, but only access to the output of the model $f$.
\textbf{Transferability.} The transferability of adversarial examples is first explored by \cite{papernot2016transferability}, indicating that an adversarial sample that causes the misclassification of a model $f'$, is also misclassified by model $f$.
To leverage this transferability against the target model, the attacker can first build a surrogate model $f'$ which conducts the same task as the target model $f$.
Then a surrogate dataset $\mathcal{D}'$ is collected and used to train the model $f'$ by querying the remote model $f$.
Finally, the attacker can craft the attacks against $f'$, and exploit the generated adversarial samples to transfer to the target model $f$.
In our work, we study the transferability of adversarial examples to surpass the $\mathcal{R}$ module of D-CAPTCHA system.
This reduces the knowledge required for attack success as in fact, we only have access to labels of the system (real or fake).

\section{Methodology}\label{sec:method}

\begin{table}[!ht]
\centering
\caption{Notation Summaration}
\label{tab:notation}
\begin{tabular}{c|c}
\textbf{Notation}                  & \textbf{Meaning}                                \\ \hline
$\mathcal{D} = (x_i, y_i)^n_{i=1}$ & Training data with $n$ is the training set size \\
$x_{adv}$                          & Adversarial sample                              \\
$\delta$                           & Perturbation                                    \\
$\mathcal{F}$                      & Target deepfake detector                        \\
$\hat{\mathcal{F}}$                & Surrogate deepfake detector                     \\
$\mathcal{C}$                      & Task classifier                                 \\
$\mathcal{I}$                      & Identify model                                  \\
$\mathcal{L}$                      & Loss function            \\
$\mathcal{V}$                      & Voice conversion model                          \\ \hline
\end{tabular}
\end{table}

\textbf{Notation.} Table \ref{tab:notation} summarizes our notations used to describe our method.

\subsection{Threat model}
Our threat model is an integration of human (adversary), the voice conversion model, and theadversarial sample generation technique, to evade the D-CAPTCHA system.
Before giving an overview of our attack, we point out the characteristics of our threat model.
Table \ref{tab:threat-model} summarizes the characteristics of our threat model.
\textbf{Attacker's Goal.} The attacker’s goal is defined based on the desired security violations, including integrity and availability violations. 
Our attack aims to evade detection of the D-CAPTCHA system without compromising normal system operation.
\textbf{Attacker's Knowledge.} The research adopts a black-box threat model where the adversary has knowledge of only the task performed by modules $\mathcal{F}$, $\mathcal{C}$, $\mathcal{I}$, and the decision output of the D-CAPTCHA system.
This means that the adversary does not have any information about training data, preprocessing techniques, feature extractors, learning algorithms with the loss function and its parameters, and inference API, in the case of Machine Learning as a Service.
Note that, in this work, we assume that all information related to the number and the types of challenges is available to the public.
\textbf{Attacker's Capability.} This characteristic defines how the system can be affected by the attacker, and how data can be manipulated.
In this case, the adversary can only manipulate the test data.
\textbf{Attacker's Strategy.} To bypass three modules $\mathcal{T}, \mathcal{R}$ and $\mathcal{C}$, the attacker trains a surrogate model by querying collected data to the target model.
The attacker then uses adversarial samples generated by it to attack the target model.
\textbf{Attack Overview.} In this threat model, the attacker's role involves understanding the challenge assigned by the D-CAPTCHA system.
To successfully surpass three modules of the system, the attacker must achieve two objectives: (i) Understanding the challenges assigned by the system; (ii) Having the generated audio adversarial sample must bypass the $\mathcal{R}$ and $\mathcal{C}$ modules.
Given our hypothesis of an imperceptible adversarial example capable of fooling the $\mathcal{C}$ module, the adversary's task is to manipulate the $\mathcal{R}$ module into classifying the generated audio sample into a human-like audio sample.
This means that after utilizing a voice conversion model to convert the audio sample's identity into a target's identity to fool the victim, the adversary would understand the provided challenges and select the corresponding adversarial samples prepared by the surrogate model to send to the system.
Mathematically, given a voice sample converted by $\mathcal{V}$ denoted as $\mathcal{V}(x)$, the attack's objective aims to find an adversarial sample $x_{adv} = \mathcal{V}(x) + \delta$ such that:
\begin{equation}
    \mathcal{F}(x_{adv}) = y, ||\delta|| < \epsilon
\end{equation}
, where $y$ is the target label; and $\epsilon$ controls the power density of $\delta$.
In our case, the target label is \textit{Real} for \textit{Fake} audio samples.

\subsection{Surrogate model: Generate imperceptible adversarial examples}\label{subset:adversarial-algo}
Attacking the surrogate model can be cast as a white-box evasion attack, where the optimization problem given in Eq. (\ref{eq:adversarial}) can be rewritten as:
\begin{equation}
    \begin{array}{cc} 
        \text{min} & \mathcal{L}_{net}(\hat{\mathcal{F}}(\mathcal{V}(x) + \delta), y) + \alpha \cdot \mathcal{L}_{\theta}(\mathcal{V}(x), \delta) \\
        \text{such that} & ||\delta|| < \epsilon 
    \end{array}
    \label{eq:optimization}
\end{equation}
, where $\alpha$ is a balance parameter.
To ensure the imperceptibility of generated adversarial examples, we utilize the frequency masking technique proposed by \citeauthor{qin2019imperceptible} \cite{qin2019imperceptible}.
The fundamental concept is to identify a masking threshold for each louder signal considered as the masker, where any signals below this threshold become inaudible to the human auditory system.
During the generation of adversarial examples, two values need to be determined: the global masking threshold of the original audio sample $\theta_x$ and the normalized log-magnitude power spectral density (PSD) estimate of the perturbation $\overline{p}_{\delta}$.
While the calculation of $\theta_x$ follows the method outlined by \citeauthor{lin2015principles} \cite{lin2015principles}, $\overline{p}_{\delta}$ can be computed via:
\begin{equation}
    \overline{p}_{\delta} = 96 - \text{max}\{p_x\} + p_{\delta}
\end{equation}
, where $p_x$ and $p_{\delta}$ are the PSD estimates of the original audio input and the perturbation, respectively.
If $\overline{p}_{\delta}$ is under $\theta_x$, the perturbation will be masked out by the original audio input and therefore be imperceptible to humans.
\textbf{Optimization.} Two objectives need to be optimized in Eq. \ref{eq:optimization}: $\mathcal{L}_{net}$ ensures the generated adversarial examples can mislead $\mathcal{F}$ module into making the desired target label $y$, while $\mathcal{L}_{\theta}$ restricts the normalized PSD estimation of the perturbation $\overline{p}_{\delta}$, ensuring that it remains below the frequency masking threshold of the original audio $\theta_x$.
This optimization can be achieved in two stages.
The first stage is to optimize $\mathcal{L}_{net}$ by clipping the perturbation to be within a small range on each iteration:

\begin{equation}
    \delta \leftarrow \text{clip}_{\epsilon}(\delta - \text{lr}_1 \cdot \text{sign}(\nabla_{\delta}\mathcal{L}_{net}(\hat{\mathcal{F}}(\mathcal{V}(x) + \delta), y)))
\end{equation}
, where $\text{lr}_1$ is the learning rate used in the first stage, $\nabla_{\delta}\mathcal{L}_{net}$ is the gradient of $\mathcal{L}_{net}$ with respect to $\delta$.
In the second stage, imperceptible adversarial samples are generated by minimizing the perceptibility via:

\begin{equation}
    \delta \leftarrow \delta - \text{lr}_2 \cdot \nabla_{\delta}[\mathcal{L}_{net}(\hat{\mathcal{F}}(\mathcal{V}(x) + \delta), y) + \alpha \cdot \mathcal{L}_{\theta}(\mathcal{V}(x), \delta)]
\end{equation}
, where $ \text{lr}_2$ is the learning rate used in the second stage.

\subsection{Black-box: Transferibility}
Previous works have explored the transferability of adversarial examples across machine learning models \cite{demontis2019adversarial, papernot2016transferability}.
These researches indicated that an adversarial example crafted to deceive one model can potentially mislead other models trained for the same task.
This is because different models trained for the same task have a substantial overlap in the error spaces, creating a significant intersection of vulnerability.
Mathematically, given an adversarial sample $x_{adv}$ optimized by the attack algorithm against the surrogate model $\hat{\mathcal{F}}$, its transferability can be defined as the loss attained by the target model $\mathcal{F}$, $T=\mathcal{L}(\mathcal{F}(x_{adv}, y))$.

Furthermore, it has been shown in \cite{demontis2019adversarial} that adversarial examples with higher confidence are more likely to transfer successfully to the target model.
In light of this finding, our objective in this work is to caft adversarial examples that induce misclassification with maximum confidence in the surrogate model $\hat{\mathcal{F}}$.

\subsection{Black-box: Imperceptible adversarial examples to Task module}
The module $\mathcal{C}$ can be cast as a binary classification problem, with the output $1$ referring to the input containing the provided task content; otherwise for the output $0$. 
To surpass this module, we hypothesize that an imperceptible adversarial sample can preserve the content of an audio sample.
Therefore, our objective is that $\mathcal{C}(x_{adv}) = 1$, where $x_{adv}$ is the imperceptible adversarial sample.

\section{Experiments}\label{sec:experiments}
\subsection{Experimental setups}\label{subsec:setup}
Note that there is no deployment of the D-CAPTCHA system or public resources provided by authors \cite{yasur2023deepfake}.
To ensure methodological consistency, each module of this system was re-implemented as described in \cite{yasur2023deepfake}.
Therefore, in this section, we present the process of building not only the surrogate model but also audio deepfake detectors and task classifiers.

We performed all our experiments on a machine with a CPU AMD EPYC 9654P, 2 GPUs NVIDIA RTX 4090, and 1492GB of RAM. 

\subsubsection{Datasets} We first describe datasets used to construct audio deepfake detectors, and then those utilized for building task classifiers.

\textbf{Audio Deepfake Datasets.} We evaluate audio deepfake detectors on three datasets, including WaveFake \cite{frank2021wavefake}, ASVspoof 2019 \cite{todisco2019asvspoof}, and ASVspoof 2021 \cite{yamagishi2021asvspoof}.
\begin{itemize}
    \item WaveFake: The dataset includes $104,885$ synthetic audios generated by $7$ generating neural networks (details in \cite{frank2021wavefake}) using $18,100$ bonafide audios from LJSpeech and JSUT datasets.
    \item ASVspoof 2019: This is the third edition in a series of challenges in audio spoofing detection, which is divided into two different use case scenarios: logical access (LA) and physical access (PA). In our case, we use the LA subset, which consists of $12,483$ bonafide and $108,978$ fake audio samples. Those synthetic samples are created using TTS and VC models.
    \item ASVspoof 2021: Similar to ASVspoof 2019, the dataset is the fourth edition which incorporates an additional task: deepfake speech detection (DF). In our case, we utilize the DF subset, encompassing $22,617$ bonafide and $589,212$ fake audio samples.
\end{itemize}
\begin{table}
\centering
\caption{The number of samples in each task dataset}
\label{tab:task_dataset}
\begin{tblr}{
  column{1} = {c},
  cell{1}{1} = {r=2}{},
  cell{1}{2} = {r=2}{},
  cell{1}{3} = {c=3}{c},
  cell{3}{1} = {r=2}{},
  cell{5}{1} = {r=2}{},
  cell{7}{1} = {r=2}{},
  cell{9}{1} = {r=2}{},
  cell{11}{1} = {r=2}{},
  vlines,
  hline{1,3,5,7,9,11,13} = {-}{},
  hline{2} = {3-5}{},
  hline{4,6,8,10,12} = {2-5}{},
}
\textbf{ Task}          & \textbf{ Dataset}                                      & \textbf{Samples } &              &               \\
                        &                                                        & \textbf{Train}    & \textbf{Val} & \textbf{Test} \\
Sing (S)                & AudioSet \cite{gemmeke2017audio}   & 2075              & 543          & 1234          \\
                        & HumTrans \cite{liu2023humtrans}       & 13080             & 765          & 769           \\
Hum Tone (HT)           & GTZAN \cite{sturm2013gtzan}           & 680               & 120          & 200           \\
                        & VocalSet \cite{wilkins2018vocalset}       & 1242              & 219          & 365           \\
Speak with Emotion (SE) & CREMA-D \cite{cao2014crema}           & 5061              & 893          & 1488          \\
                        & RAVDESS \cite{livingstone2018ryerson} & 998               & 172          & 288           \\
Laugh (L)               & AudioSet \cite{gemmeke2017audio}      & 943               & 166          & 277           \\
                        & VocalSound \cite{gong2022vocalsound}  & 2278              & 526          & 700           \\
Domestic Sound (DS)     & AudioSet \cite{gemmeke2017audio}      & 385               & 67           & 112           \\
                        & DASED \cite{serizel2020sound}         & 992               & 176          & 692           
\end{tblr}
\end{table}

\textbf{Task Datasets.} To evaluate our hypothesis regarding imperceptible adversarial examples, we employ three tasks similar to those presented in \cite{yasur2023deepfake}, including \textbf{S}ing a song, \textbf{H}um \textbf{T}une for a song, \textbf{S}peak with \textbf{E}motion. 
Additionally, we also introduce two tasks: create \textbf{D}omestic \textbf{S}ound, and \textbf{L}augh.
However, regarding the AudioSet datasets for these two tasks, we only select audio samples that belong to both the speech and laugh classes.
This diverse selection of tasks enables us to comprehensively evaluate the effectiveness of our approach in generating imperceptible adversarial examples across a range of audio manipulation tasks.
Table \ref{tab:task_dataset} summarizes the number of train, val, and test samples divided in each task datasets.
Each task utilizes two publicly available datasets. 
If any dataset provides pre-defined train, validation, and test splits, we will employ those splits. 
Otherwise, we will randomly split each dataset into ratios of 65\%, 15\%, and 20\% for train, validation, and test sets, respectively.
Note that in terms of the DESED dataset, we only use the validation recorded soundscapes subset for our study since it is labeled, then split it into 85\% for the training subset, and 15\% for the validation subset, while the public YouTube recorded soundscapes leveraged for the testing subset.

\subsubsection{Voice Conversion}
We select three current voice conversion models to evaluate the inference time and intelligibility: kNN-VC \cite{baas2023voice}, TriAAN-VC \cite{park2023triaan}, and Urhythmic \cite{van2023rhythm}.
We leverage pre-trained models provided by those authors and evaluate their inference performance on the VCTK dataset \cite{Yam2019VCTK}.
Specifically, we randomly select audio recordings from the VCTK dataset with durations from $7$ to $15$ seconds, allowing for a comprehensive analysis across varying audio lengths.
Each generated sample has a sample rate of $16,000$ Hz.
Regarding intelligibility, we calculate the word/character error (W/CER) between the target audio and the converted one.

\subsubsection{Audio Deepfake Detectors} 
In this section, we present the procedure for constructing the surrogate model and target models.
Table \ref{tab:train-deepfake} shows the performance of trained deepfake detectors, including surrogate model and target models.

\begin{table}[htbp]
\centering
\caption{Training Performance of Audio Deepfake Detectors}
\label{tab:train-deepfake}
\begin{tblr}{
  cells = {c},
  cell{3}{1} = {r=3}{},
  vline{2} = {1-2, 3-5}{},
  hline{2-3,6} = {-}{},
}
                & \textbf{Model} & \textbf{Precision} & \textbf{Recall} & \textbf{F1 Score} \\
Surrogate model & LCNN           & 0.981              & 0.987           & 0.984             \\
Target model    & SpecRNet       & 0.945              & 0.992           & 0.968             \\
                & RawNet2        & 0.985              & 0.981           & 0.972             \\
                & RawNet3        & 0.994              & 0.977           & 0.985             
\end{tblr}
\end{table}

\textbf{Surrogate model.}

As demonstrated by \citeauthor{demontis2019adversarial} \cite{demontis2019adversarial}, the adversarial sample generated by a low-complexity surrogate model can highly succeed against both low and high-complexity target models.
This means that a low-complexity model is less vulnerable to adversarial attacks than its high-complexity counterpart.
Therefore, in this work, we select LCNN \cite{wu2020light} as our surrogate model for creating adversarial audio samples.
We utilize the WaveFake dataset for training this model with linear frequency cepstral coefﬁcients (LFCC) frontend.
%
As previously mentioned, we assume that information regarding the types and number of tasks is publicly accessible; thus, the attacker also collects task datasets for building the surrogate model.
This implies that there exists an adversarial audio sample for each task.
In this work, we use the VocalSet dataset for the Hum Tone task, the CREMA-D dataset for the Speak with Emotion task, and the AudioSet datasets for the Sing, Laugh, and Domestic Sound tasks.
Moreover, to address the class imbalance in the dataset, we employ undersampling to reduce the number of fake samples to match the number of benign samples.
After collecting the necessary datasets, those audio samples are labeled by querying to target models.
LCNN is trained for $5$ epochs, batch size $128$, with Adam optimizer, and binary focal loss \cite{lin2017focal}.

\textbf{Target models.}
RawNet2 \cite{tak2021end}, SpecRNet \cite{kawa2022specrnet}, RawNet3 \cite{kawa2022defense} are three deepfake detectors employed in our experiments.
Due to the unavailability of pre-trained models, we re-implemented these models using ASVspoof 2019 and ASVspoof 2021 datasets.
Hyperparameters and configurations are set following the descriptions provided in their respective papers.
For each detection method, we train the model with $25$ epochs, a batch size of $128$, and report the test result achieved at the epoch corresponding to the model's best performance on the validation set.

\subsubsection{Task Classifiers}
We re-implement $\mathcal{C}$ module by constructing GMM \cite{sahidullah2015comparison}, ResNet18 \cite{he2016deep}, and RawNet3 \cite{kawa2022defense} models.
Table \ref{tab:train-task} presents the training performance of these classifiers on each task.

\begin{table}[htbp]
\centering
\caption{Training Performance of Task Classifiers.}
\label{tab:train-task}
\begin{tblr}{
  cells = {c},
  cell{2}{1} = {r=3}{},
  cell{5}{1} = {r=3}{},
  cell{8}{1} = {r=3}{},
  cell{11}{1} = {r=3}{},
  cell{14}{1} = {r=3}{},
  vline{2-5} = {1-4, 5-7, 8-10,11-13,14-16}{},
  vline{3-5} = {3-4,6-7,9-10,12-13,15-16}{},
  hline{2,5,8,11,14,17} = {-}{},
}
\textbf{Dataset} & \textbf{Models} & \textbf{Precision} & \textbf{Recall} & \textbf{F1 score} \\
HumTrans         & GMM             & 0.912              & 0.937           & 0.961             \\
                 & ResNet18        & 0.973              & 0.996           & 0.968             \\
                 & RawNet3         & 0.994              & 0.979           & 0.986             \\
GTZAN            & GMM             & 0.892              & 0.737           & 0.842             \\
                 & ResNet18        & 0.924              & 0.657           & 0.896             \\
                 & RawNet3         & 0.928              & 0.841           & 0.881             \\
RAVDESS          & GMM             & 0.952              & 0.931           & 0.956             \\
                 & ResNet18        & 0.979              & 0.768           & 0.861             \\
                 & RawNet3         & 0.986              & 0.989           & 0.987             \\
VocalSound       & GMM             & 0.898              & 0.904           & 0.932             \\
                 & ResNet18        & 0.973              & 0.996           & 0.965             \\
                 & RawNet3         & 0.957              & 0.922           & 0.965             \\
DASED            & GMM             & 0.805              & 0.824           & 0.701             \\
                 & ResNet18        & 0.832              & 0.602           & 0.724             \\
                 & RawNet3         & 0.882              & 0.861           & 0.811             
\end{tblr}
\end{table}

The purpose of selecting these three models is to investigate the influence of different preprocessing techniques on imperceptible adversarial samples.
Specifically, for each classifier, preprocessing techniques are applied and hyperparameters are set up as follows:
\begin{itemize}
    \item GMM: Mel-frequency cepstral coefficients (MFCC) is utilized to model features of audio signal, with parameters: length of the analysis window is $0.05$, the step between successive windows is $0.02$, the number of cepstrum is $10$, and the Fast Fourier Transform (FFT) size is $800$.
    \item ResNet18: Since Resnet is the model for image classification tasks, we need to convert audio samples into spectrogram-based features. We utilize default Spectrogram function of \textit{torchaudio} package, with changed parameters: $\text{n\_fft} = 2048$ and $\text{hop\_length} = 512$.
    \item RawNet3: This is a speaker recognition model that directly operates on raw waveform inputs; thus no preprocessing technique is used in this model.
\end{itemize}
We use HumTrans, GTZAN, RAVDESS, VocalSound, and DASED datasets for training those models, which means that each task has three corresponding datasets.
Except for GMM, we train each classifier for $5$ epochs, a batch size of $128$, Cross Entropy loss, and Adam optimizer.

\begin{algorithm}
  \DontPrintSemicolon
  \SetKwInOut{Input}{Input}
  \Input{Model parameter $\theta_0$, number of PGD steps $t$, minibatch $B$}

  \For{(x, y) in B}{
    Let $x_0 = x$ \;
    \For{$i = 0, \cdot, t - 1$}{
        $x_{i+1} \leftarrow \text{Proj}_{\Delta(x)}(x_{i} + \alpha \cdot \text{sign}(\nabla_x \mathcal{L}(\mathcal{F}_{\theta}(x_{i}), y)))$
    }
    Update $\theta \leftarrow \theta - \beta \nabla_{\theta}\mathcal{L}(\mathcal{F}_{\theta}(x_t), y)$
  }
  \caption{Adversarial training used for task classifiers}
  \label{alg:advtraining}
\end{algorithm}

\textbf{Adversarial Training.} We improve the robustness of task
classifiers by utilizing the Algorithm \ref{alg:advtraining}. We use the same
datasets defined above to train ResNet and RawNet with the
same number of PGD steps t = 20 and t = 40.

\subsection{Metrics}
\textbf{Evaluate trained models.} We employ F1-score for evaluating the effectiveness of our trained deepfake detectors and task classifiers.
The F1-score is a widely recognized metric for assessing the performance of binary classification tasks, particularly those involving imbalanced datasets. 
F1-score is calculated as the harmonic mean of recall and precision as follows:
\begin{equation*}
    F1 - score = 2 \times \frac{\text{Precision} \times \text{Recall}}{\text{Precision} + \text{Recall}}
\end{equation*}
, where $\text{Recall} = \frac{TP}{TP+FP}$ and $\text{Precision} = \frac{TP}{TP+FN}$.
In terms of VC models, WER and CER are utilized to evaluate their intelligibility.
W/CER measures the average number of words/characters that are incorrectly recognized compared to the reference transcript.
In VC models, it measures the errors between the target samples and the corresponding converted ones.

\textbf{Evaluate attacks.} We use Attack Success Rate (ASR) to measure the fraction of samples that bypass the surrogate model and target models.

\begin{table}[htbp]
\centering
\caption{Comparision of VC's Intelligibility.}
\label{tab:intelligibility}
\begin{tabular}{c|c|c}
\textbf{Model} & \textbf{WER (\%)} & \textbf{CER (\%)}  \\ 
\hline
kNN-VC         & 25.78             & 15.67              \\ 
\hline
Urhythmic      & 37.12             & 24.68              \\ 
\hline
TriAAN-VC      & 19.87             & 11.25              \\
\hline
\end{tabular}
\end{table}

\subsection{Results}
\subsubsection{Qualification on Voice Conversion models}
\begin{figure}[htbp]
    \centering
     \includegraphics[width=.85\linewidth]{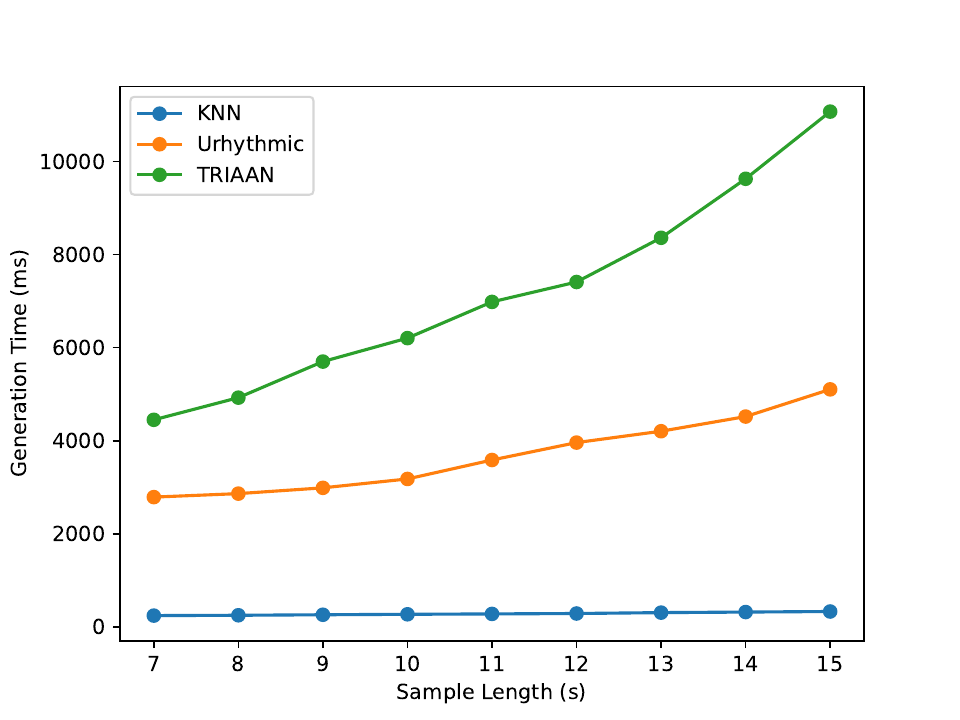}
    \caption{Comparision of VC's Inference Speed.}
    \label{fig:VC-speed}
\end{figure}

We examine the inference speed of three recently introduced voice conversion models: kNN-VC, TriAAN-VC, and Urhythmic.
Our selection of these models specifically targets the most recent advancements in voice conversion technology.
The table \ref{tab:intelligibility} and figure \ref{fig:VC-speed} show the fast inference speed and generation quality of kNN-VC.
This evaluation indicates that kNN-VC satisfies the stringent requirement of the D-CAPTCHA system, generating a synthetic audio sample in a single second while maintaining a high level of understandability.
Therefore, kNN-VC serves as our voice conversion of choice for fooling the victim.
It is noteworthy that the adversary continues to employ voice conversion for subsequent communications with the victim even after bypassing the D-CAPTCHA; thus using it is necessary.

\begin{table}[htbp]
\centering
\caption{Attack Success Rate ($\%$) of Transferability from surrogate model to target deepfake detectors}
\label{tab:transfer-deepfake}
\begin{tblr}{
  cells = {c},
  cell{1}{1} = {r=2}{},
  cell{1}{2} = {c=4}{},
  vline{2} = {1-2}{},
  vline{3-5} = {2}{},
  vline{2-5} = {3}{},
  hline{2} = {2-5}{},
  hline{3} = {-}{},
}
Surrogate model  & Transferability on Target Models (\%) &                   &                  &                  \\
                 & \textbf{LCNN}                         & \textbf{SpecRNet} & \textbf{RawNet2} & \textbf{RawNet3} \\
\textbf{LCNN}    & 99.76                                 & 41.87             & 35.91            & 36.83           
\end{tblr}
\end{table}

\subsubsection{Evaluation on Transferibility}
In this section, we conduct two main experiments to evaluate the transferability capability of the surrogate model to target models and to examine our hypothesis about imperceptible adversarial examples that might bypass the task classifiers.
In the first experiment, we only use a subset of $13,421$ fake samples from WaveFake's test subset to transfer to three target models: SpecRNet, RawNet2, and RawNet3.
The purpose is to test transferability to deepfake detectors because most of these models are only trained with datasets that only include speech.
This subset includes two types of adversarial audio samples: high-confidence and low-confidence samples, allowing us to investigate whether high-confidence adversarial samples exhibit better transferability compared to their low-confidence counterparts.
From the table \ref{tab:transfer-deepfake}, we can observe that:
\begin{enumerate}
    \item The success rate of transferring adversarial samples from LCNN to SpecRNet is higher than RawNet2 and RawNet3. This is because of the effect of feature extraction techniques as LCNN and SpecRNet employ the same front-end technique (LFCC) while RawNet2 and RawNet3 directly operate on raw waveforms. This observation suggests that feature extraction techniques hinder the transferability of adversarial examples across audio deepfake detection.
    \item Majority of successfully transferable samples have high confidence, indicating that higher-confidence adversarial attacks have a greater likelihood of successfully deceiving target deepfake detectors.
\end{enumerate}

In the second experiment, we examine the transferability of imperceptible adversarial examples to three target task classifiers: GMM, ResNet, and RawNet.
This investigation aims to address two key questions: (i) Will adversarial samples involving specific contents, such as a song, influence the transfer success rate to deepfake detectors? (ii) Will the transferability of imperceptible adversarial examples be impacted by feature extraction techniques employed by these task classifiers?

\begin{table}[htbp]
\centering
\caption{Attack Success Rate (\%) of Transferability}
\label{tab:transfer}
\begin{tabular}{c|c|c|c}
\textbf{Task} & \textbf{Deepfake Detectors} & \multicolumn{2}{c}{\textbf{Task Classifiers}}  \\

\hline
              & \textbf{SpecRNet}                & \textbf{ResNet18} & \textbf{RawNet3}    \\ 
\hline
S             & 37.16                            & 32.57           & 34.28              \\
HT            & 35.93                            & 30.16           & 34.58              \\
SE            & 38.58                            & 36.41           & 37.68              \\
L             & 32.04                            & 26.14           & 28.71              \\
DS            & 29.76                            & 24.75           & 27.83              \\ 
\hline
              & \textbf{RawNet2}                 & \textbf{ResNet18} & \textbf{RawNet3}    \\ 
\hline
S             & 33.97                            & 28.96           & 31.68              \\
HT            & 31.45                            & 27.01           & 29.96              \\
SE            & 34.83                            & 32.68           & 34.12              \\
L             & 29.12                            & 24.60           & 27.11              \\
DS            & 25.73                            & 23.38           & 25.65              \\ 
\hline
              & \textbf{RawNet3}                 & \textbf{ResNet18} & \textbf{RawNet3}    \\ 
\hline
S             & 33.83                            & 30.06           & 32.17              \\
HT            & 30.86                            & 28.79           & 30.48              \\
SE            & 35.25                            & 33.61           & 34.01              \\
L             & 29.05                            & 25.56           & 28.26              \\
DS            & 26.41                            & 22.31           & 25.12              \\
\hline
\end{tabular}
\end{table}

To address our above questions, we use $6,451$ synthetic task-specific samples and then pass them to each deepfake detector and task classifier.
For example, considering the singing task, we select corresponding perturbed samples from $6,451$ samples.
These examples are then sequentially fed into SpecRNet.
Upon successfully evading the detection by SpecRNet, those samples are further evaluated using a task classifier GMM.
This approach allows us to thoroughly investigate the transferability of adversarial examples across both deepfake detection and task classification systems, while also examining the impact of specific content embedded in the adversarial samples on their effectiveness.
From table \ref{tab:transfer}, we can observe that:
\begin{enumerate}
    \item None of the task-specific adversarial samples can transfer successfully to deepfake detectors. The attack success rate decreases significantly for each deepfake detector, especially for DS task. This can be explained by the complexities of retaining specific audio features, such as sounds like "closing/opening the door", when adding perturbations. 
    \item The success rate of RawNet is higher than GMM and ResNet. This is because those sounds are also not robust to feature extraction techniques of task classifiers. As mentioned earlier, feature extraction techniques used for GMM and ResNet are MFCC and spectrogram-based, respectively while RawNet operates directly on raw waveform.
\end{enumerate}

\subsubsection{Evaluate on Robustness of Task Classifiers}
\begin{figure*}[!ht]
\centering
\begin{subfigure}[b]{.9\linewidth}
  \centering
  \includegraphics[width=.9\linewidth]{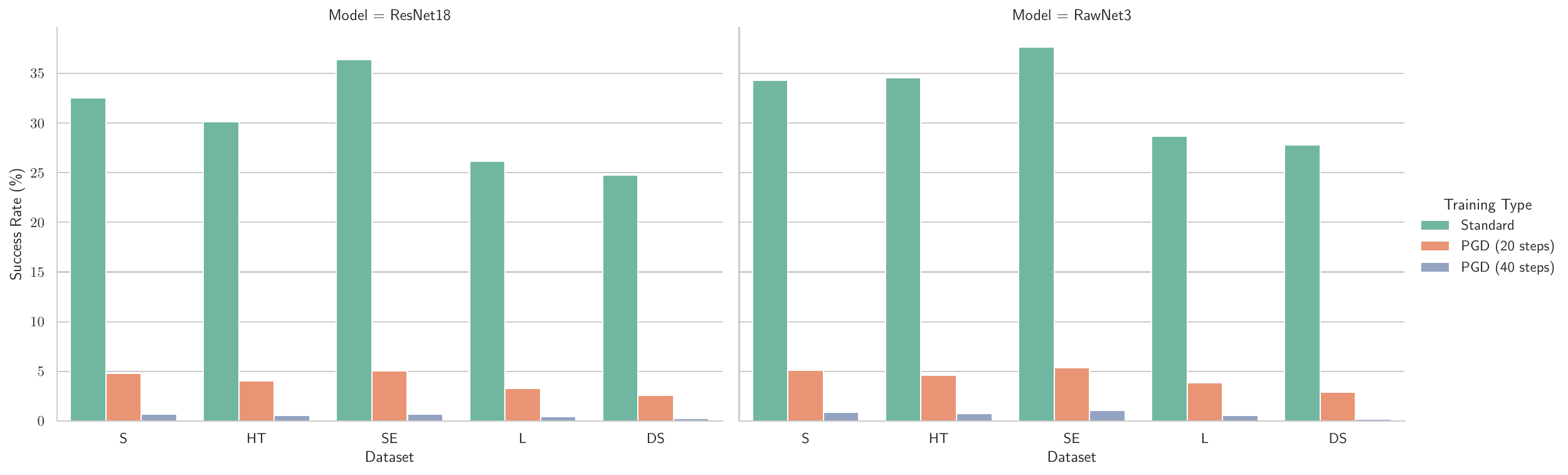}
  \caption{Task Classifiers}
  \label{fig:task-classifier}
\end{subfigure}
\\
\begin{subfigure}[b]{.9\linewidth}
  \centering
  \includegraphics[width=.9\linewidth]{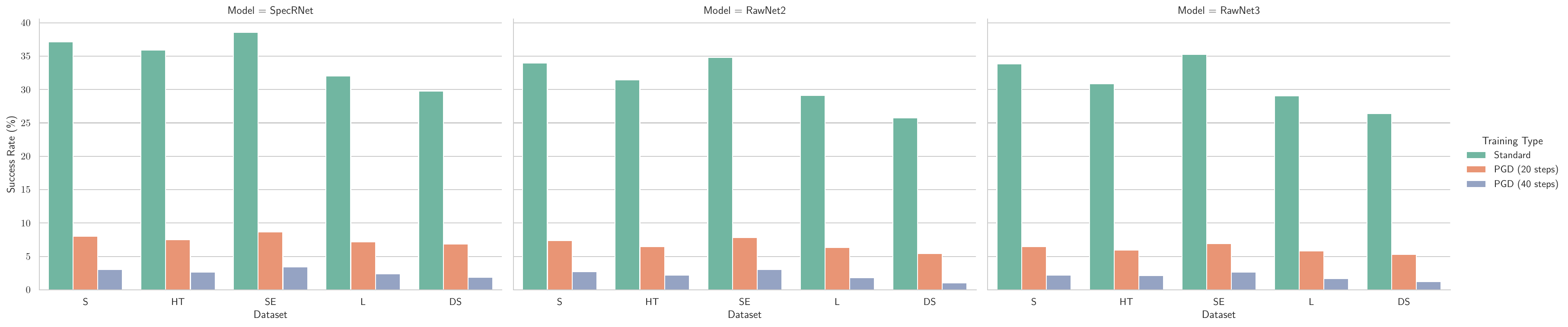}
  \caption{Deepfake Detectors}
  \label{fig:deepfake-detector}
\end{subfigure}
\caption{Attack Success Rate of Task classifiers and Deepfake detectors before and after applying PGD adversarial training.}
\label{fig:adversarial-training}
\end{figure*}

In this experiment, we evaluate the performance of deepfake detectors and task classifiers after employing PGD adversarial training.
Table \ref{tab:compare} compares the ASR of D-CAPTCHA and D-CAPTCHA++ while figure \ref{fig:adversarial-training} shows the changes in ASR after applying PGD adversarial training.
From table \ref{tab:compare} and figure \ref{fig:adversarial-training}, we can observe that:
\begin{itemize}
    \item The ASR decreases significantly on both deepfake detectors and task classifiers of D-CAPTCHA++. Particularly, the ASR of deepfake detectors and task classifiers reduces from $32.26 \% \pm 0.99$ to $2.27 \% \pm 0.18$ and from $31.31 \% \pm 1.40$ to $0.60 \% \pm 0.09$, respectively.
    \item When the number of PGD steps increases to $t=40$, the ASR can significantly decline to nearly $0\%$, presenting the effectiveness of adversarial training in improving the robustness of both deepfake detectors and task classifiers against adversarial samples.
\end{itemize}

\begin{table*}[!ht]
\centering
\fontsize{6pt}{6pt}\selectfont
\caption{Attack Success Rate ($\%$) of D-CAPTCHA and D-CAPTCHA++}
\label{tab:compare}
\centering
\begin{tblr}{
  cells = {c},
  cell{1}{1} = {r=3}{},
  cell{1}{2} = {c=3}{},
  cell{1}{5} = {c=6}{},
  cell{2}{2} = {c=3}{},
  cell{2}{5} = {c=3}{},
  cell{2}{8} = {c=3}{},
  cell{3}{3} = {c=2}{},
  cell{3}{6} = {c=2}{},
  cell{3}{9} = {c=2}{},
  vline{2-3} = {1-3}{},
  vline{3,6} = {2}{},
  vline{3-4,6-7,9} = {3}{},
  vline{2-3,5-6,8-9} = {1-21}{},
  hline{2-4} = {2-10}{},
  hline{5,10-11,16-17,22} = {-}{},
}
\textbf{Task } & \textbf{D-CAPTCHA }         &                           &         & \textbf{D-CAPTCHA++ }             &                            &         &                                   &                            &         \\
               & \textbf{Standard Training}  &                           &         & \textbf{PGD Training (20 steps) } &                            &         & \textbf{PGD Training (40 steps) } &                            &         \\
               & \textbf{Deepfake Detectors} & \textbf{Task Classifiers} &         & \textbf{Deepfake Detectors}       & \textbf{Task Classifiers } &         & \textbf{Deepfake Detectors}       & \textbf{Task Classifiers } &         \\
               & SpecRNet                    & ResNet18                  & RawNet3 & SpecRNet                          & ResNet18                   & RawNet3 & SpecRNet                          & ResNet18                   & RawNet3 \\
S              & 37.16                       & 32.57                     & 34.28   & 8.03                              & 4.77                       & 5.13    & 3.06                              & 0.67                       & 0.91    \\
HT             & 35.93                       & 30.16                     & 34.58   & 7.47                              & 4.05                       & 4.64    & 2.62                              & 0.58                       & 0.77    \\
SE             & 38.58                       & 36.41                     & 37.68   & 8.64                              & 5.08                       & 5.34    & 3.45                              & 0.81                       & 1.05    \\
L              & 32.04                       & 26.14                     & 28.71   & 7.21                              & 3.31                       & 3.88    & 2.37                              & 0.41                       & 0.54    \\
DS             & 29.76                       & 24.75                     & 27.83   & 6.87                              & 2.56                       & 2.91    & 1.85                              & 0.21                       & 0.38    \\
               & RawNet2                     & ResNet18                  & RawNet3 & RawNet2                           & ResNet18                   & RawNet3 & RawNet2                           & ResNet18                   & RawNet3 \\
S              & 33.97                       & 28.96                     & 31.68   & 7.38                              & 4.37                       & 4.91    & 2.74                              & 0.59                       & 0.74    \\
HT             & 31.45                       & 27.01                     & 29.96   & 6.48                              & 3.77                       & 4.17    & 2.17                              & 0.52                       & 0.64    \\
SE             & 34.83                       & 32.68                     & 34.12   & 7.81                              & 4.86                       & 5.19    & 3.05                              & 0.73                       & 0.93    \\
L              & 29.12                       & 24.60                     & 27.11   & 6.36                              & 3.11                       & 3.45    & 1.81                              & 0.35                       & 0.41    \\
DS             & 25.73                       & 23.38                     & 25.65   & 5.45                              & 2.64                       & 2.76    & 1.03                              & 0.16                       & 0.22    \\
               & RawNet3                     & ResNet18                  & RawNet3 & RawNet3                           & ResNet18                   & RawNet3 & RawNet3                           & ResNet18                   & RawNet3 \\
S              & 33.83                       & 30.06                     & 32.17   & 6.46                              & 3.66                       & 3.81    & 2.21                              & 0.49                       & 0.67    \\
HT             & 30.86                       & 28.79                     & 30.48   & 5.97                              & 3.17                       & 3.58    & 2.12                              & 0.42                       & 0.57    \\
SE             & 35.25                       & 33.61                     & 34.01   & 6.93                              & 4.44                       & 4.81    & 2.68                              & 0.71                       & 0.88    \\
L              & 29.05                       & 25.56                     & 28.26   & 5.81                              & 2.31                       & 2.49    & 1.65                              & 0.28                       & 0.32    \\
DS             & 26.41                       & 22.31                     & 25.12   & 5.32                              & 1.63                       & 2.01    & 1.25                              & 0.06                       & 0.17    
\end{tblr}
\end{table*}

\section{Discussion}\label{sec:discussion}

In this paper, we evaluate the resilience of the D-CAPTCHA
system in a black-box manner in which attackers may only
query the target model and obtain the system’s final output.
Prior works also evaluate different automatic speech/ speaker
recognition under black-box settings but require more than
200, 000 queries to generate adversarial samples successfully \cite{taori2019targeted, wang2020towards}. In this work, our surrogate model can be built to generate imperceptible adversarial samples by using $51,270$ queries. Specifically, we first construct a surrogate model to generate imperceptible adversarial examples, and then utilize them to transfer to target models. Moreover, we demonstrate our hypothesis that the imperceptibility of adversarial audio samples can bypass the task classifiers, and also indicate the vulnerability of the D-CAPTCHA system to adversarial examples. Therefore, we propose to apply the PGD adversarial training method to deepfake detectors and task classifiers to enhance the robustness of
the D-CAPTCHA system.

Based on our evaluation results, we have several recommendations for designing defense methods against adversarial
attacks:
\begin{itemize}
    \item \textbf{Adversarial training should be applied for task classifiers}. Adversarial samples should be created and involvedin the training set, helping to improve the generalization
and robustness of the classifiers.
    \item \textbf{Imbalanced datasets should be considered.} When constructing a detection-based defense, it is meaningless if it detects most bonafide audio samples as adversarial. This is mostly caused by the imbalance in the training dataset where the number of fake samples is more than the natural ones. Therefore, we suggest reporting the evaluation results with different metrics, not only the accuracy but also the F1-score, and ROC curve.
    \item \textbf{Feature extraction should be applied for voice-based
    DNN.} Our experimental results indicate that deepfake detectors and task classifiers employing feature extraction techniques (e.g., MFCC, spectrogram) have less vulnerability to transferable adversarial samples.
\end{itemize}
However, there are some limitations in our research: (i) The
generation of imperceptible adversarial examples cannot be
fully guaranteed against the Identity module I because the
introduction of perturbations into audio samples might lead to
discrepancies in identity between $a_0$ and $r_c$ ; (ii) We have not
evaluated our attack over the telephony network that might
cause the loss of perturbations during transmission because of
codec compression, and static interference.

Our future work will focus on three areas: (i) Extend
our proposed attack to guarantee the identity similarity when
adding perturbations to audio samples; (ii) Study the robust-
ness of imperceptible adversarial samples over the air and over
telephony network.

\bibliography{template}
\end{document}